\documentclass[referee]{raa} 
\usepackage{natbib}
\usepackage{graphicx,times}   
\usepackage{amssymb,amsmath}
\usepackage{multirow}
\usepackage[colorlinks=true, linkcolor=blue, citecolor=blue, urlcolor=blue]{hyperref}

\begin{document}

   \title{Identifications of RR Lyrae stars and Quasars from the simulated data of Mephisto-W Survey
}

   \volnopage{Vol.0 (20xx) No.0, 000--000}      
   \setcounter{page}{1}          
   \author{Lei Lei \inst{1} 
    \and Bing-Qiu Chen \inst{2} 
    \and Jin-Da Li \inst{1} 
    \and Jin-Tai Wu \inst{1} 
    \and Si-Yi Jiang \inst{1} 
    \and Xiao-Wei Liu \inst{2} 
   }

   \institute{Department of Astronomy, Yunnan University,
   Kunming, Yunnan 650500, China; \\
        \and
        South-Western Institute for Astronomy Research, Yunnan University, 
        Kunming, Yunnan 650500, China; {\it bchen@ynu.edu.cn}\\
\vs\no
   {\small Received~~20xx month day; accepted~~20xx~~month day}}

\abstract{We have investigated the feasibilities and accuracies of the identifications of RR Lyrae stars and quasars from the simulated data of the 
Multi-channel Photometric Survey Telescope (Mephisto) W Survey. Based on the variable sources light curve libraries from the Sloan Digital Sky Survey (SDSS) 
Stripe 82 data and the observation history simulation from the Mephisto-W Survey Scheduler, we have simulated the $uvgriz$ multi-band light curves of RR Lyrae stars,
quasars and other variable sources for the first year observation of Mephisto W Survey. We have applied the  
ensemble machine learning algorithm Random Forest Classifier (RFC) to identify RR Lyrae stars and quasars, respectively. 
We build training and test samples and extract $\sim$ 150 features from the simulated light curves and train two RFCs respectively for the 
RR Lyrae star and quasar classification. We find that, our RFCs are able to select the RR Lyrae stars and quasars with remarkably high 
precision and completeness, with $purity$ = 95.4\,per\,cent and $completeness$ = 96.9\,per\,cent for the RR Lyrae RFC and 
$purity$ = 91.4\,per\,cent and $completeness$ = 90.2\,per\,cent for the quasar RFC. We have also derived relative importances of the extracted features utilized 
to classify RR Lyrae stars and quasars.
\keywords{methods: data analysis ---
surveys ---
catalogs ---
stars: variables: RR Lyrae ---
quasars: general
}
}
   \authorrunning{Lei et al. }            
   \titlerunning{Identification of Mephisto RR Lyrae stars and Quasars}  
   \maketitle
%
%
%
\section{Introduction}           
\label{sect:intro}

The Multi-channel Photometric Survey Telescope (Mephisto; \citealt{Yuan+etal+2020}) is a 
wide-field survey telescope with a 1.6\,m primary mirror. Mephisto has a field of view of $\sim$ 2.36\,deg$^2$.
It is equipped with three CCD cameras and is capable of imaging the same patch of sky in three bands simultaneously.  
The telescope will be installed at
Lijiang Observatory in the Southwest of China before the end of 2021.
During 2022 and 2031, Mephisto will carry out a ten-year survey program which have two components: the Mephisto-W survey 
and the Mephisto-D, H and M surveys (Er et al. in prep.). All the observing time of the first year of the survey (2022) 
will be allocated to the Mephisto-W survey. The full survey area (the northern sky of $\sim$ 27,000\,deg$^2$ of declination
between $-$21\degr\ and 75\degr) will be imaged several times in both the $ugi$ and 
$vrz$ filter combinations over the year, using pairs of 20-second exposures (\citealt{Lei+etal+2021}; Chen et al. submitted). 
Two key science goals of the Mephisto-W Survey are the Galactic archeology,
and the studies of the distant galaxies and cosmology. 
The RR Lyrae variable stars are important tracers for the study of the Milky Way (e.g. \citealt{Sesar+etal+2010};
\citealt{Ablimit+etal+2017}; \citealt{Ablimit+etal+2018};  \citealt{Liu+etal+2020}; \citealt{Hattori+etal+2020}; \citealt{Griv+etal+2020}; \citealt{Andres+etal+2021}; \citealt{Ablimit+etal+2021}). 
Large sample of quasars will allow us to probe the nature of them (e.g. \citealt{Kuo+etal+2012}; \citealt{Pasquet+etal+2018}) and to constrain the cosmological parameters (e.g. \citealt{Khadka+etal+2021}; \citealt{Mediavilla+etal+2021}). 
Thus to identify the RR Lyrae stars and quasars from the data of the Mephisto-W survey and to obtain complete and un-contaminated samples of them are fundamental to achieve those key science goals of the Mephisto-W survey. 

Chen et al. (submitted) have presented the the Mephsito-W
Survey Scheduler (MWSS) and provide the simulations of the first year observations of the Mephisto-W Survey.
In the current work, we have simulated the Mephisto-W survey observations of variable objects, including the RR Lyare
stars, quasars and other variable sources, based on
Chen et al. simulation and the light curve libraries of variable objects from the literature. 
We have trained Random Forest Classifiers (RFCs) to identify RR Lyrae stars and 
quasars respectively from the simulated data of Mephisto-W Survey and obtained the 
 accuracies and completeness of the classifiers.
 
In Sect.~\ref{sect:Data}, we introduce how we simulate the observations of different variable objects
of the Mephisto-W survey. In Sect.~\ref{sect:ML} we describe the RFCs we adopted to identify RR Lyrae stars and quasars.
In Sect.~\ref{sect:discussion} we show our results, which are discussed and summarized in Sect.~\ref{sect:conclusion}.

\section{Simulated data}
\label{sect:Data}

The process of the realizations of the Mephisto W Survey observed RR Lyrae stars, quasars and other variable sources
includes two steps: the simulation of the observing cadence of the Mephisto-W Survey 
and that of the light curves of the individual variable sources.

For the cadence simulation, we adopt the Simulation~1 from Chen et al. (submitted) in the current work. 
Chen et al. (submitted) have presented an adaptive scheduling algorithm for the Mephisto-W Survey. 
The scheduler can simulate the observational results of the Mephito-W survey with giving
models of the telescope, weather conditions and other environmental variables.
Chen et al. have provided two sets of simulation results for the first year observation of the Mephisto-W Survey.
In the current work, we adopt the first simulation, i.e., Simulation~1 from Chen et al.
For Simulation~1, 48.1\,per\,cent and 30.7\,per\,cent of the survey fields 
would be targeted by the Mephisto respectively in the $ugi$ and $vrz$ filter combinations for more than five times.
In the current work, we focus on the Sloan Digital Sky Survey (SDSS; \citealt{York2000}) Stripe 82 region,
where most of the fields will be targeted by the Mephisto five times in a year for both the $ugi$ and $vrz$ 
filter combinations.

\begin{figure}
\centering
\includegraphics[width=\textwidth, angle=0]{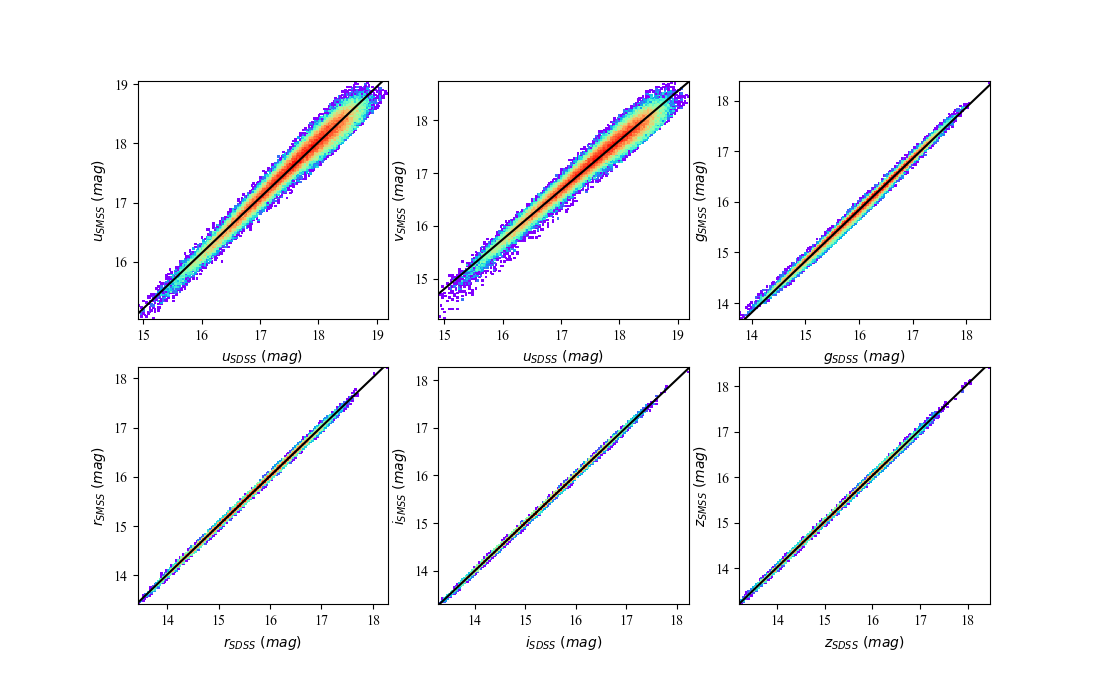}
\caption{Relationships between the SkyMapper $uvgriz$ magnitudes and the SDSS $ugriz$ magnitudes 
for the individual stars in the SDSS Stripe 82 Standard Star Catalog. The black lines show the 
best-fit linear relations. }
\label{Fig2}
\end{figure}

We have adopted a method similar as that of  \citet{Oluseyi+etal+2012} to simulate the Mephisto-W Survey 
observations of variable sources.  To assess the capability of characterizing RR Lyrae stars from the Legacy Survey of Space and Time (LSST), 
\citet{Oluseyi+etal+2012} have undertaken extensive simulations of RR Lyrae star light curves from the LSST operation simulations
and the SDSS Stripe 82 photometric measurements. In the current work,
the simulations are also based on the SDSS Stripe 82 observations.  
\citet{Ivezic+etal+2007} have provided SDSS $ugriz$ light curves of 67,507 variable sources in the 
SDSS Stripe 82 region, including 483 RR Lyrae stars (\citealt{Sesar+etal+2010}; \citealt{Suveges+etal+2012}), 9,258 quasars \citep{Palanque-Delabrouille+etal+2011}, and 57,766 
other variable sources. All objects have an average of ten observations in each of the $ugriz$ passbands.

The filter set of the Mephisto
includes six $uvgriz$ passbands, which are very similar to that of the SkyMapper \citep{Bessell2011,Wolf2018}. As the Mephisto filters are still under developing,
in the current work, we simply adopt the SkyMapper $uvgriz$ bands as the Mephisto filters. We first transform the SDSS $ugriz$ photometric magnitudes
to the SkyMapper $uvgriz$ magnitudes. We cross-match the SDSS Stripe 82 Standard Star Catalog \citep{Ivezic+etal+2007} to the SkyMapper 
Southern Survey Data Release 2 (SMSS DR2; \citealt{Onken+etal+2019}).
In Fig.~\ref{Fig2}, we show the correlations between the SkyMapper and SDSS magnitudes.
The SkyMapper $u,~v,~g,~r,~i$ and $z$ magnitudes are simply converted from the SDSS $u,~u,~g,~r,~i$ and $z$ respectively by linear transforming relationships, as,
  \begin{eqnarray}
   u_{\rm SMSS}=0.936\ast u_{\rm SDSS}+1.183, \\
   v_{\rm SMSS}=0.938\ast u_{\rm SDSS}+0.739, \\
   g_{\rm SMSS}=1.012\ast g_{\rm SDSS}-0.353, \\
   r_{\rm SMSS}=1.001\ast r_{\rm SDSS}+0.009, \\
   i_{\rm SMSS}=1.003\ast i_{\rm SDSS}-0.045, \\
   z_{\rm SMSS}=1.004\ast z_{\rm SDSS}-0.036.
 \label{eq:transf}
 \end{eqnarray}
Based on the above equations, we are then able to obtain the idealized Mephisto $uvgriz$ light curves of the RR Lyrae stars, quasars and other variable sources 
from their SDSS $ugriz$ light curves and finally produce the Mephisto ``observed'' light curves of the individual objects.

\begin{figure}
   \centering
   \includegraphics[width=\textwidth, angle=0]{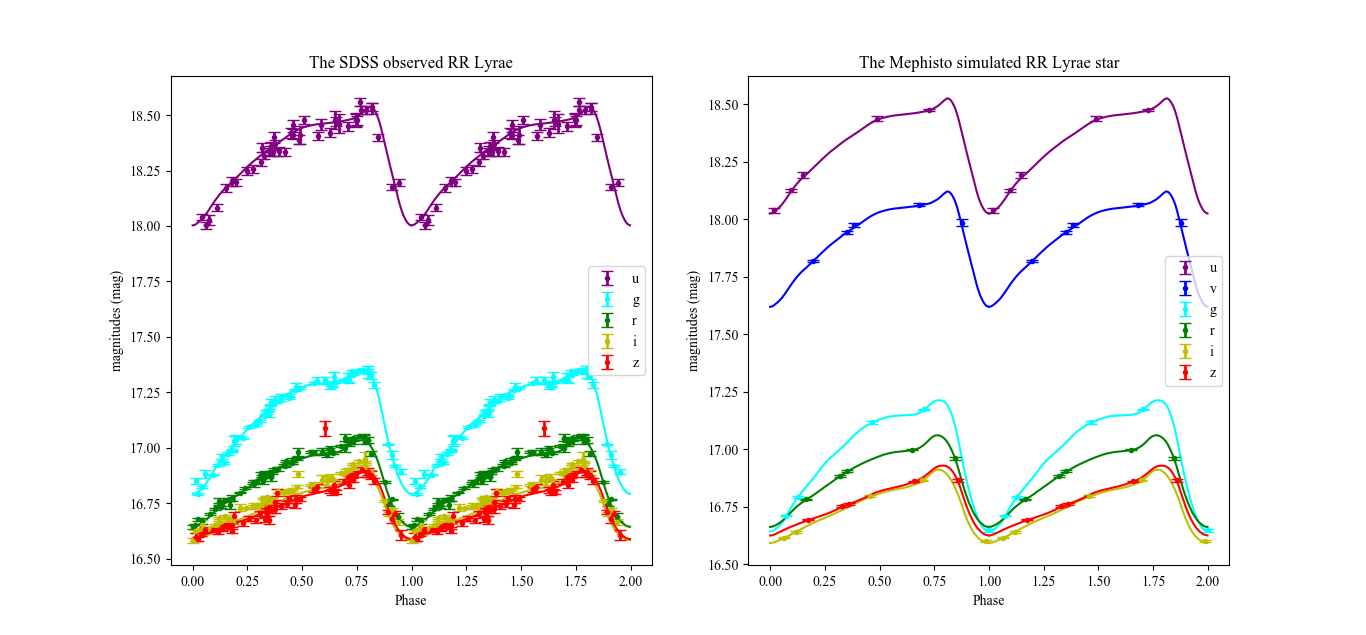}
   \includegraphics[width=\textwidth, angle=0]{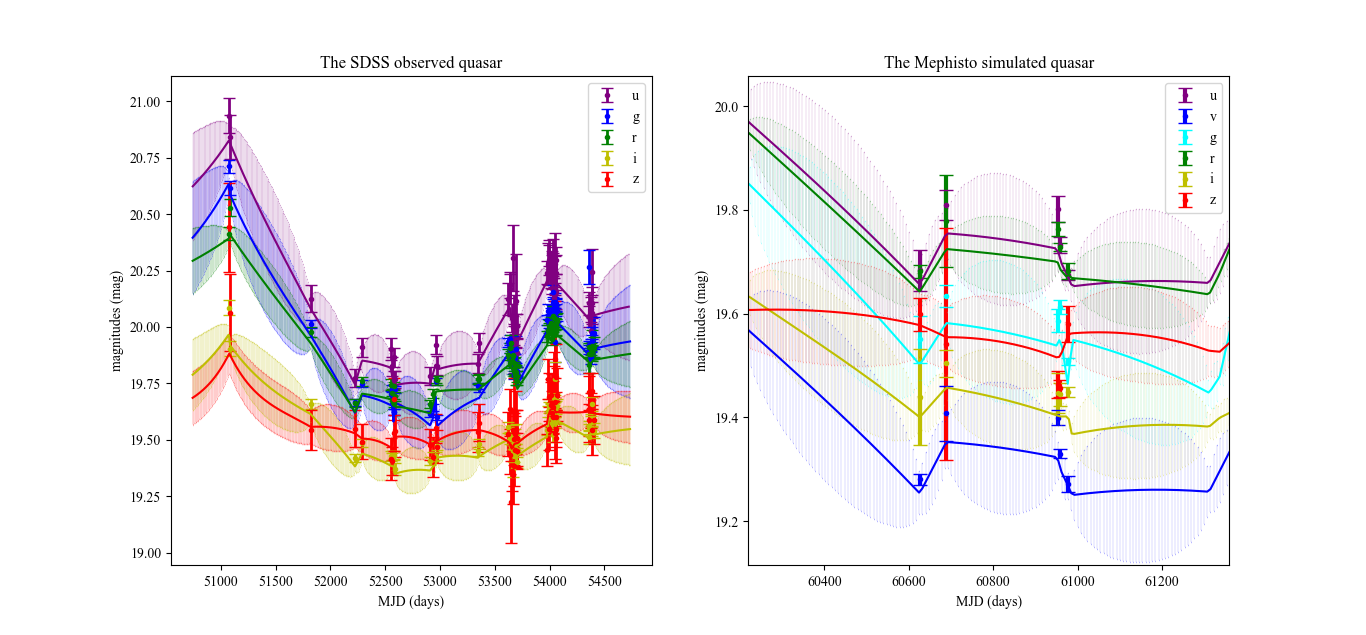}
   \caption{Examples of simulated light curves for a periodic object (RR Lyrae star; upper panels) and a non-periodic object 
   (quasar; bottom panels). For the periodic object, its observed (left) and simulated (right) light curves are plotted as functions of
   phase. For the non-periodic object, its observed (left) and simulated (right) light curves are plotted as functions of modified Julian dates. 
   For the quasar, we also show the best-fitted DRW models.}
   \label{Fig3}
   \end{figure}

The cadence simulation from Chen et al. (submitted) provides us the observing time and the observing conditions of the fields in the SDSS Strip 82
for the first year observation of the Mephisto W Survey. For the periodic 
objects such as the RR Lyrae stars, Cepheids and eclipsing binaries, etc., we calculated their phases $\phi$ at the individual epochs based on their periods $P$ and
the start time of each period $\phi_0$. We then derived the idealized magnitudes of the individual objects at each epoch based on linear interpolation of 
their phase-folded light curves. To produce realistic observations, random Gaussian noises are added to the idealized magnitudes based on the 
photometric errors calculated from the observing conditions (\citealt{Lei+etal+2021}).

For the non-periodic objects, such as the quasars, we are not able to predict their magnitudes at given epochs. We thus
randomly selected five SDSS observations which were
taken within one calendar year and manually changed their observing time to the same time of the same day of the year 2022. 
Similar as the periodic objects, the Gaussian random noises were added.
In Fig.~\ref{Fig3} we show two examples of the simulated light curves in the Mephisto $uvgriz$ bands for both the periodic and non-periodic objects.

\section{Classfication Algorithms}
\label{sect:ML}

We use a machine learning algorithm, the Random Forest Classifier (RFC; \citealt{Breiman+etal+2001}), 
to identify the RR Lyrae stars and quasars in the current work. 
RFC is a ensemble learning method for classification which fits a number of decision tree classifiers and uses all the weak classifiers collaboratively 
to improve the predictive accuracy and control over-fitting. 
The {\sc scikit-learn} package for {\sc{python}}
(\citealt{Pedregosa+etal+2011}) is adopted to build the RFCs in the current work.
Based on the simulated Mephisto light curves of different variable sources, we have built separate
RFC models for identifying the RR Lyrae stars and quasars, respectively.
For the identification of RR Lyrae stars, the sample containing all the 483 RR Lyrae stars in the SDSS Strip 82 region 
\citep{Sesar+etal+2010} is adopted as the positive sample and a sample containing 483 randomly selected non-RR Lyrae stars
from \citet{Ivezic+etal+2007} is adopted as the negative sample.
For the identification of quasars, a sample containing 
9,107\footnote{The \citet{Palanque-Delabrouille+etal+2011} catalog contains 9,258 quasars, among which 9,107 have more than five visits during a calendar year.} 
quasars is adopted as the positive 
sample and a sample containing 9,107 randomly selected non-quasars is adopted as the negative sample. 

\subsection{Training features}

The simulated light curves of the objects in the positive and negative samples have been transformed into 
sets of features, which are adopted as the input parameters of the RFC models. We adopt different
sets of training features for the RR Lyrae star and quasar RFC models, respectively.

\subsubsection{RR Lyrae star RFC training features}

\citet{Vicedomini+etal+2020} have transfered the  
LSST simulated light curves into a set of features
that represent the peculiar characteristics of the variables.
With the extracted features as input parameters, \citet{Vicedomini+etal+2020} have carried out 
several Machine Learning algorithms to identify different types of supernovae.
In the current work, we adopted all the statistical parameters from 
\citet{Vicedomini+etal+2020} which are listed as follows for the RR Lyrae star RFC.
\begin{itemize}
\item Amplitude (ampl): half of the difference between the maximum and the minimum magnitudes.
\item Beyond1std (b1std): the fraction of observations that have magnitudes outside the 1$\sigma$ range from the mean value.
\item Flux percentage ratio (fpr): the ratio between two flux percentiles $F_{n,m}$, where
$F_{n, m}$ is the difference between the flux values at $n$th and $m$th percentiles, respectively.
In the current work, we adopt five flux percentile ratios:
$fpr20 = F_{40,60}/F_{5,95}$,
$fpr35 = F_{32.5,67.5}/F_{5,95}$,
$fpr50 = F_{25,75}/F_{5,95}$,
$fpr65 = F_{17.5,82.5}/F_{5,95}$, and 
$fpr80 = F_{10,90}/F_{5,95}$.
\item Lomb-Scargle periodogram (ls): the period from the
Lomb-Scargle periodogram.  For the identification of RR Lyrae stars, we adopted period limits from 0.2 to 1.2\,day. We note that for both the RR Lyrae stars and 
quasars, we are not likely to obtain the true periods of the objects. This is because that we have only simulated measurements at four to five epochs.  
\item Linear trend (lt): the slope of the light curve by a linear fit.
\item Median absolute deviation (mad): the median value of the fluxes deviated from the median value.
\item Median Buffer range percentage (mbrp): the fraction of observations that have magnitudes with 10\% from the median value.
\item Magnitude ratio (mr):  the fraction of observations that have magnitudes above the median value.
\item Maximum slope (ms): the maximum value of slopes calculated from the observations at successive epochs.
\item Percent difference flux percentile (pdfp): the ratio between the difference of the fifth and the 95th percentile flux (converted to magnitudes), and the median magnitude.
\item Pair slope trend (pst): the fraction of flux measurements that are larger than the former ones in the last 30 couples of consecutive observations.
\item R Cor Bor (rcb): the fraction of observations that have magnitudes below 1.5\,mag with respect to the median value.
\item Small Kurtosis (kurt): the fourth-order momentum divided by the square of the variance.
\item Skewness (skew):  the third-order momentum divided by the variance to the third power. 
\item Standard deviation (std): the standard deviation of the observed fluxes.
\end{itemize}
We have light curves of objects in six $uvgriz$ passbands, which resulted 114 input features as the Vicedomini et al statistical parameters
for each RR Lyrae star or non RR Lyrae star.

In addition to the Vicedomini et al statistical parameters, we have
also adopted the statistical parameters listed as follows.
\begin{itemize}
\item Colors (color): the colors derived from the average magnitude values of two bands, $color = \overline{mag_i} - \overline{mag_j}$, where $i$ and $j$ are
the indices of the filters. In this paper we have adopted 15 colors: $\bar{u}-\bar{v},~\bar{u}-\bar{g},~\bar{u}-\bar{r},~\bar{u}-\bar{i},~\bar{u}-\bar{z},~\bar{v}-\bar{g},~\bar{v}-\bar{r},~\bar{v}-\bar{i},~\bar{v}-\bar{z},~\bar{g}-\bar{r},~\bar{g}-\bar{i},~\bar{g}-\bar{z},~\bar{r}-\bar{i},~\bar{r}-\bar{z}$ and $\bar{i}-\bar{z}$.
\item Mean values of the real-time colors (mrcolor): the average values of the real-time colors, $mrcolor = \overline{rcolor}$, where $rcolor$ is the real-time color. 
For the Mephisto W Survey, the observations are 
made in either the $ugi$ or $vrz$ filter combinations. Thus we would have six real-time colors: $u-g,~u-i,~g-i,~v-r,~v-z$ and $r-z$, for which the magnitudes are
obtained at the same time.
 \item Amplitudes of the real-time colors (ampcolor): the differences between the maximum and the minimum values of the real-time colors, 
 $ampcolor = rcolor_{\rm max} - rcolor_{\rm min}$.
\end{itemize}
In total, we have adopted 141 input parameters for the RR Lyrae star RFC.

\subsubsection{Quasar RFC training features}

For the quasar RFC model, we used also all the 141 parameters adopted by the RR Lyrae star RFC.
In addition, similar as in the works of \citet{MacLeod2010} and \citet{Yang+etal+2021}, we have adopted the Damped Random Walk (DRW) parameters,
including the time scale of DRW $\tau$ and the long-term deviation of variability $\sigma$,
as the input features of the quasar RFC model. 
The {\sc{javelin}} program is adopted to fit the light curves in each passbands 
to calculate the DRW parameters $\tau$ and $\sigma$ \citep{Zu+etal+2013}, which 
resulted 12 additional input features. 

\section{Experiment}
\label{sect:discussion}

The performances of the RR Lyrae star and quasar RFCs are based on some statistical estimators. For a given class (i.e., RR Lyrae star or quasar), we define 
$True Positive$ as the number of objects which are correctly classified as the class;
$False Positive$ as the number of objects which are wrongly classified as the class, 
but their correct classifications are not the class; 
$True Negative$ as the number of objects which are correctly classified as not the class, 
and $False Negative$ as the number of objects which are wrongly classified as not the class, 
but their correct classification are the class. 
We then have:
\begin{eqnarray}
Purity=\dfrac{TruePositive}{TruePositive + FalsePositive}, \\
Completeness =\dfrac{TruePositive}{TruePositive + FalseNegative}.
\end{eqnarray}
$Purity$ of the RFC model is also named as $precision$. 
It is the percentage of that a certain type of classification is true.
$Completeness$ of the RFC model is also named as $recall$. 
It is the percentage of the correctly classified objects for a given class of objects.

We divided both the positive and negative samples into the same number of subsets. 
Each time, we select some of the subsets for RFC model training and the remaining subsets for testing the trained classifiers.
The values of $purity$ and $completeness$ of each classifier are recorded and finally 
we present the averaged performances.

\subsection{Performance of the RR Lyrae star identification}

\begin{table}
   \begin{center}
   \caption[]{ The averaged values of $purity$ and $completeness$ of the RR Lyrae Star RFC.}
   \label{Tab:RRcp}
\begin{tabular}{ccc}
   \hline\noalign{\smallskip}
 ~ & RR Lyrae star & Non RR Lyrae star \\
   \hline\noalign{\smallskip}
 Average $purity$  & 0.954 & 0.969     \\
 Average $completeness$  & 0.969 & 0.953     \\
   \noalign{\smallskip}\hline
 \end{tabular}
 \end{center}
 \end{table}
 
The RR Lyrae star positive and negative samples contains both 483 objects. They are divided into 48 subsets, which are noted as $S1$, $S2$, $S3$, ..., 
$S47$ and $S48$. The last subset ($S48$) contains 13 RR Lyrae stars and 13 non RR Lyrae objects; and the other 47 subsets 
all contain 10 RR Lyrae stars and 10 non RR Lyrae objects.
We train the RR Lyrae star RFC model 48 times. At each time, 36 subsets are selected as the training sample and the other 12 subsets as test sample.
For example, at the first time, the subsets $S1$, $S2$, $S3$, ..., $S35$ and $S36$ are adopted as the training sample and the remaining subsets 
($S37$, $S38$, $S39$, ..., $S47$ and $S48$) as the test sample. 
At the second time, the subsets $S2$, $S3$, $S4$, ...,  $S36$ and $S37$ are adopted as the training sample and the  remaining subsets
($S38$, $S39$, $S40$, ..., $S48$ and $S1$) as the test sample.

We present the averaged performance of our RR Lyrae classifiers in Table~\ref{Tab:RRcp}.
We find a high performance of our RR Lyrae RFC. The precision of RR Lyrae star classification can achieve  95.4\,per\,cent and 
the recall 96.9\,per\,cent, which clearly demonstrates the high efficiency of selecting RR Lyrae star from the
data of the Mephisto-W Survey. 

 \begin{figure}
   \centering
   \includegraphics[width=\textwidth, angle=0]{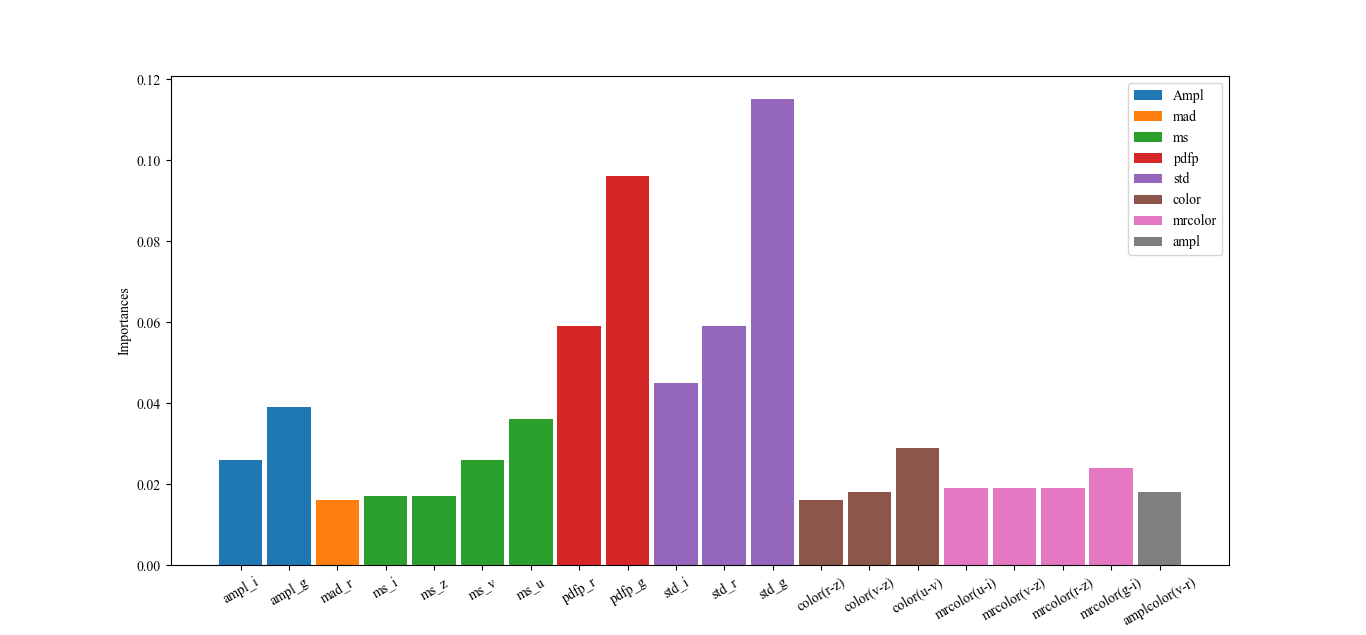}
   \includegraphics[width=\textwidth, angle=0]{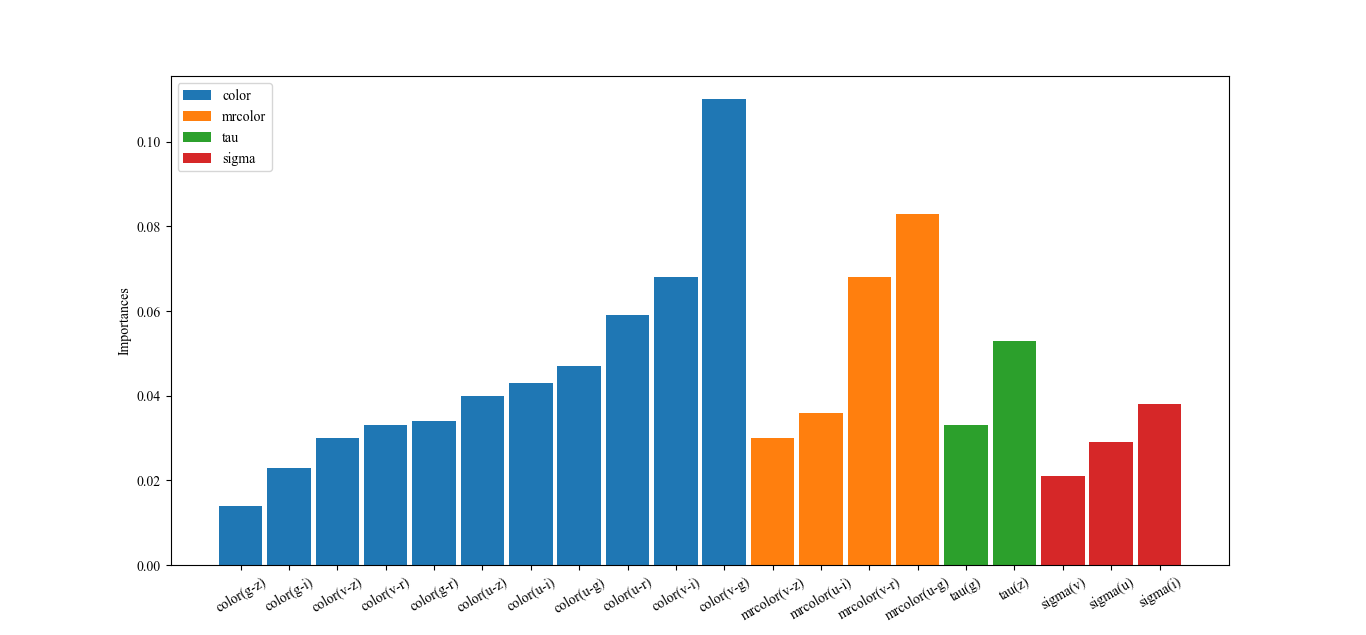}
   \caption{Important scores of 20 most important input parameters for the RR Lyrae star (upper) and the 
   quasar (bottom) RFCs, respectively.}
   \label{important}
   \end{figure}

For the RR Lyrae star RFC, we have adopted 141 input features for classifier training. We have examined the 
relative importance of these input features. Since we have trained the RR Lyrae star RFC 48 times, for each trial,
we also record the important score of every input feature. We show the averaged scores of 20 most important features in the 
upper panel of Fig.~\ref{important}. The most important features are standard deviations (std),  percent difference flux percentiles (pdfp),
amplitudes (ampl), maximum slopes (ms), colors (color) and mean values of the real-time colors (mrcolor).
In particular the std and pdfp in $g$-band are two most important features. 

\subsection{Performance of the quasar identification}

\begin{table}
   \begin{center}
   \caption[]{ The average values of $purity$ and $completness$ of the quasar RFC.}
   \label{Tab:QSOcp}
\begin{tabular}{ccc}
   \hline\noalign{\smallskip}
 ~ & Quasar & Non quasar \\
   \hline\noalign{\smallskip}
 Average $purity$  & 0.914 & 0.903     \\
 Average $completness$  & 0.902 & 0.915     \\
   \noalign{\smallskip}\hline
 \end{tabular}
 \end{center}
 \end{table}

The quasar positive and negative samples contains both 9,107 objects. They are divided into 91 subsets. The last subset contains 107 quasars
and 107 non quasars, while the other subsets contain 100 quasars and 100 non quasars. 
Similar as the training of the RR Lyrae star RFC, we have trained the quasar RFC 91 times. 
At each time, 68 subsets are adopted as the training sample and the remaining 23 subsets as the test sample.
We present the averaged performance of the quasar classifiers in Table~\ref{Tab:QSOcp}.  
The precision of quasar classification is 91.4\,per\,cent and 
the recall 90.2\,per\,cent. The performance of the quasar classifiers are not as good as the RR Lyrae star classifiers.
However, it is still possible for us to select the quasar candidates from the Mephisto-W Survey for the considerably high precision and recall.

We have also examined the relative importance of the input features for the quasar RFC, which 
is presented in the bottom panel of Fig.~\ref{important}. The most important features are the colors (color),  mean values of the real-time colors (mrcolor) 
and the DRW parameters ($\tau$ and $\sigma$).
Particularly, the color $\bar{v}-\bar{g}$ and mrcolor $u-g$ are two most important features. 

\section{Discussion and conclusions}
\label{sect:conclusion}

The Mephisto-W survey will target the whole northern sky of $\sim$27,000\,deg$^2$. All the available time in the first year of the survey 
will be dedicated to Mephisto-W. The full survey area will be imaged four to five times over the year, in both the $ugi$ and $vrz$ filter combinations. 
The present work is related to the key sciences of the Mephisto-W survey, with special emphasis to the identifications of RR Lyrae stars and quasars.

In order to explore the the feasibilities and accuracies of selecting
RR Lyrae star and quasar from the first year observation of the Mephisto-W Survey, we have simulated the $uvgriz$ multiband light curves 
of the RR Lyare stars, quasars and other variable objects based on the Mephisto-W Survey Scheduler simulation and the light curve catalogs of 
the variable sources from the SDSS Stripe 82 observations. We then trained RFCs respectively for the RR Lyrae stars and quasars and investigated
the accuracies and recalls of the classifiers.

For the RR Lyrae star identification, we have built positive and negative samples containing 483 RR
Lyrae stars and 483 non RR Lyrae stars, respectively. 141 observation features were extracted from their simulated light curves and were applied to the
RR Lyrae star RFC training. We have obtained average values of 95.4 and 96.9\,per\,cent respectively 
for the $precision$ and $completeness$ of the RR Lyrae star RFC, which indicate that we are able to select RR Lyrae star
from the Mephisto W survey data with very high efficiency.
For the quasar identification, we have built positive and negative samples containing 9,107 quasars and
9,107 non quasars, respectively. 153 training features are adopted. The trained RFC can select the quasars with 
a $precision$ of 91.4\,per\,cent and a $completeness$ of 90.2\,per\,cent.

RFC adopts bagging and random feature sampling methods, which has good resistance to noise. Using the same method as \citet{Breiman+etal+2001}, we have tested the noise effect of our classifiers. We artificially set the input labels of 5\% objects in the training sample to the wrong labels. This noise injection leads to errors of ~ 0.04\% and 0.4\% for the RR Lyrae star and quasar RFCs, respectively. This indicates that the RFC method is insensitive to noises and the classifier is stable.

The Mephisto telescope is planned to obtain its first light in the end of 2021 and the Mephisto-W Survey will target the whole northern sky of ~27,000 deg2. Although the Mephisto-W survey fields would be targeted by the telescope for only four to five times in a year, we are still able to identify the RR Lyrae stars and quasars with high accuracies. This is benefited from the high accuracy real-time colors obtained by the Mephisto-W survey, Comparing to the traditional method which select RR Lyrae stars and quasars from (period) analysis of light curves of the individual objects, our machine learning algorithm takes much less time and computing resources. It will be powerful for the modern large-scale time domain surveys, which will deliver observations of billions sources. In addition, our method do not require many epochs observations, which saves the telescope time and enables us to cover much larger areas.

Our method can be applied directly to the Mephisto data once it is available. The algorithm can also be
applied to the data of other time-domain surveys, such as the Zwicky Transient Facility (ZTF; \citealt{Mahabal+etal+2019}; \citealt{Graham+etal+2019}; \citealt{Bellm+etal+2019}), Wide Field Survey Telescope (WFST; \citealt{Chen+etal+2019}; \citealt{Lou+etal+2020}), LSST and China Space Station Telescope (CSST; \citealt{Zhao+etal+2016}; \citealt{Yuan+etal+2021}; \citealt{Sun+etal+2021}; \citealt{Cao+etal+2021A}; \citealt{Cao+etal+2021C}).


\begin{acknowledgements}
This work is funded by the National Natural Science Foundation of China (NSFC)
No.~11803028, 1183300 and 12173034, National Training Program of Innovation and Entrepreneurship for 
Undergraduates of China No. 201910673001, Yunnan University grant C176220100007 and 
National Key R\&D Program of China No. 2019YFA0405500.
We acknowledge the science research grants from the China Manned Space Project 
with NO. CMS-CSST-2021-A09, CMS-CSST-2021-A08 and CMS-CSST-2021-B03.

Funding for SDSS-III has been provided by the Alfred P. Sloan Foundation, the Participating Institutions, the National Science Foundation, and the U.S. Department of Energy Office of Science. The SDSS-III web site is http://www.sdss3.org/.
SDSS-III is managed by the Astrophysical Research Consortium for the Participating Institutions of the SDSS-III Collaboration including the University of Arizona, the Brazilian Participation Group, Brookhaven National Laboratory, Carnegie Mellon University, University of Florida, the French Participation Group, the German Participation Group, Harvard University, the Instituto de Astrofisica de Canarias, the Michigan State/Notre Dame/JINA Participation Group, Johns Hopkins University, Lawrence Berkeley National Laboratory, Max Planck Institute for Astrophysics, Max Planck Institute for Extraterrestrial Physics, New Mexico State University, New York University, Ohio State University, Pennsylvania State University, University of Portsmouth, Princeton University, the Spanish Participation Group, University of Tokyo, University of Utah, Vanderbilt University, University of Virginia, University of Washington, and Yale University.

The national facility capability for SkyMapper has been funded through ARC LIEF grant LE130100104 from the Australian Research Council, awarded to the University of Sydney, the Australian National University, Swinburne University of Technology, the University of Queensland, the University of Western Australia, the University of Melbourne, Curtin University of Technology, Monash University and the Australian Astronomical Observatory. SkyMapper is owned and operated by The Australian National University's Research School of Astronomy and Astrophysics. The survey data were processed and provided by the SkyMapper Team at ANU. The SkyMapper node of the All-Sky Virtual Observatory (ASVO) is hosted at the National Computational Infrastructure (NCI). Development and support of the SkyMapper node of the ASVO has been funded in part by Astronomy Australia Limited (AAL) and the Australian Government through the Commonwealth's Education Investment Fund (EIF) and National Collaborative Research Infrastructure Strategy (NCRIS), particularly the National eResearch Collaboration Tools and Resources (NeCTAR) and the Australian National Data Service Projects (ANDS).

\end{acknowledgements}




\bibliographystyle{raa}
\bibliography{sample-cbq}
\label{lastpage}

\end{document}